\documentclass[twocolumn,aps,prl,superscriptaddress]{revtex4}
\usepackage{graphicx}
\usepackage{bm}

\def\bea{\begin{eqnarray}}
\def\eea{\end{eqnarray}}
\def\be{\begin{equation}}
\def\ee{\end{equation}}

\begin{document}

\title{Optical Production of Stable Ultracold $^{88}$Sr$_2$ Molecules}

\author{G. Reinaudi}
\affiliation{Department of Physics, Columbia University, 538 West 120th Street, New York, New York 10027-5255, USA}
\author{C. B. Osborn}
\affiliation{Department of Physics, Columbia University, 538 West 120th Street, New York, New York 10027-5255, USA}
\author{M. McDonald}
\affiliation{Department of Physics, Columbia University, 538 West 120th Street, New York, New York 10027-5255, USA}
\author{S. Kotochigova}
\affiliation{Department of Physics, Temple University, Philadelphia, Pennsylvania 19122, USA}
%\author{R. Moszynski}
%\affiliation{Quantum Chemistry Laboratory, Department of Chemistry, University of Warsaw, Pasteura 1, 02-093 Warsaw, Poland}
\author{T. Zelevinsky}
\email{tz@phys.columbia.edu}
\affiliation{Department of Physics, Columbia University, 538 West 120th Street, New York, New York 10027-5255, USA}

\begin{abstract}     %600 characters including spaces
We have produced large samples of stable ultracold $^{88}$Sr$_2$ molecules in the electronic ground state in an optical lattice.
%The molecules are bound by 0.05 cm$^{-1}$ and are stable for several milliseconds even in the presence of the surrounding $^{88}$Sr atoms.
The fast, all-optical method of molecule creation involves a near-intercombination-line photoassociation pulse followed by spontaneous emission with a near-unity Franck-Condon factor.  The detection uses excitation to a weakly bound electronically excited vibrational level corresponding to a very large dimer and yields a high-$Q$ molecular vibronic resonance.  This is the first of two steps needed to create deeply bound $^{88}$Sr$_2$ for frequency metrology and ultracold chemistry.
%in the absolute ground quantum state.  Lattice-trapped Sr$_2$ is of interest to frequency metrology and ultracold chemistry.


\end{abstract}
\date{\today}
\maketitle

\newcommand{\w}{3.25in}

\newcommand{\MolCrLossSchematic}[1][\w]{
\begin{figure}[h]
\includegraphics*[width=4.4in]{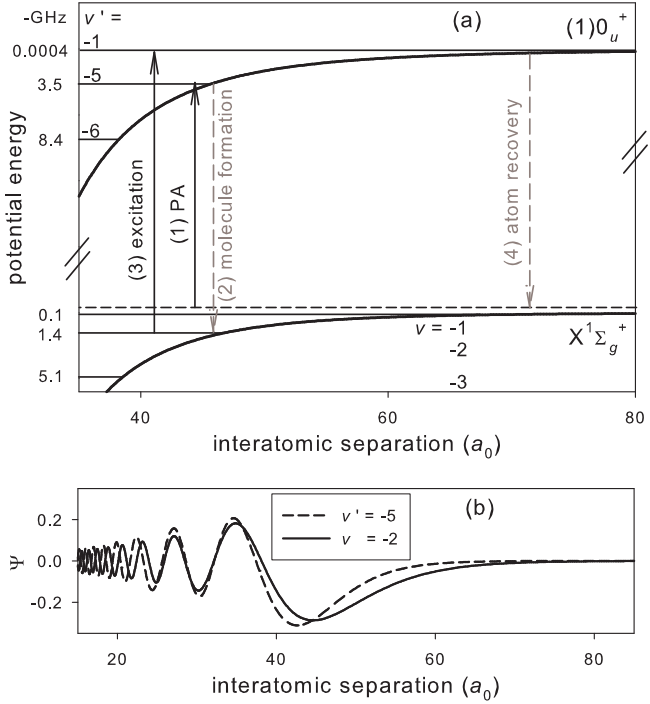}
\caption{(a) Long-range $^{88}$Sr$_2$ potential energies versus internuclear separation.  The two curves relevant to this work are the ground state potential $X^1\Sigma_g^+$ and the excited state potential $(1)0_u^+$ that dissociate to the $^1S_0+^1S_0$ and $^1S_0+^3P_1$ (689 nm) atomic limits, respectively.  Several vibrational energy levels $v,v'$, with the total molecular angular momenta $(J,J')=(0,1)$, are shown, along with their binding energies; the primed values refer to the excited potential.  The negative $v,v'$ values refer to counting from the dissociation limit, with $-1$ corresponding to the highest-lying level.  The natural line widths of the shown $v'$ levels are $\sim20$ kHz.  The horizontal dashed line represents the thermal continuum for the $\mu$K atoms.  The vertical solid arrows indicate one-photon optical pathways used for molecule creation and atom recovery.  The vertical dashed arrows indicate decay pathways completing the molecule creation and recovery; these rely on the unusually large FCF and optical length, respectively.  (b) The large value of the FCF $f_{vv'}=f_{(-2,-5)}\approx0.8$ results from an excellent agreement of the outer turning points, as illustrated by the wave function plots for these vibrational levels.}
\label{fig:MolCrLossSchematic}
\end{figure}
}

\newcommand{\FCFmeasurements}[1][\w]{
\begin{figure}[h]
\includegraphics*[width=3.5in]{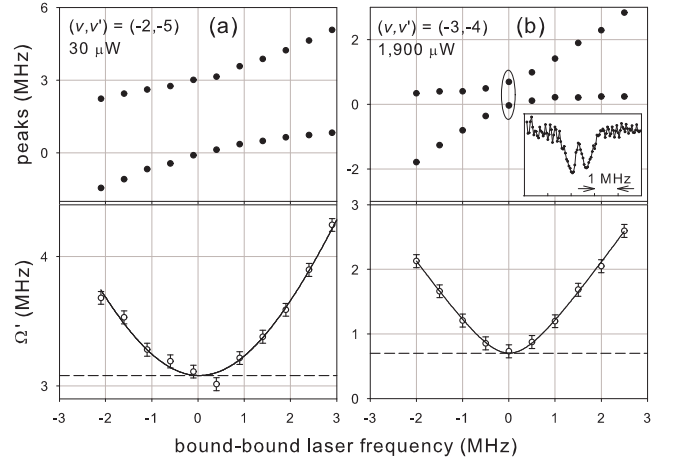}\hfill
\caption{Autler-Townes doublets emerge when the L$_{\rm{BB}}$ frequency is fixed near a molecular resonance while L$_{\rm{FB}}$ is scanned.  (a) Autler-Townes peak positions, along with their separation $\Omega'$ fitted to Eq. (\ref{eq:OmegaGen}), for $(v,v')=(-2,-5)$.  The dashed line indicates the on-resonance Rabi frequency $\Omega$ at the indicated L$_{\rm{BB}}$ power.  (b) The same as (a), for $(v,v')=(-3,-4)$.  The L$_{\rm{BB}}$ power is $60\times$ larger, while $\Omega$ is $4.4\times$ smaller, resulting in $f_{(-3,-4)}/f_{(-2,-5)}=0.9\times10^{-3}$.  The inset shows a representative atom loss curve with an Autler-Townes doublet corresponding to the circled data points.}
\label{fig:FCFmeasurements}
\end{figure}
}

\newcommand{\MolCrLoss}[1][\w]{
\begin{figure}[h]
\includegraphics*[width=3.4in]{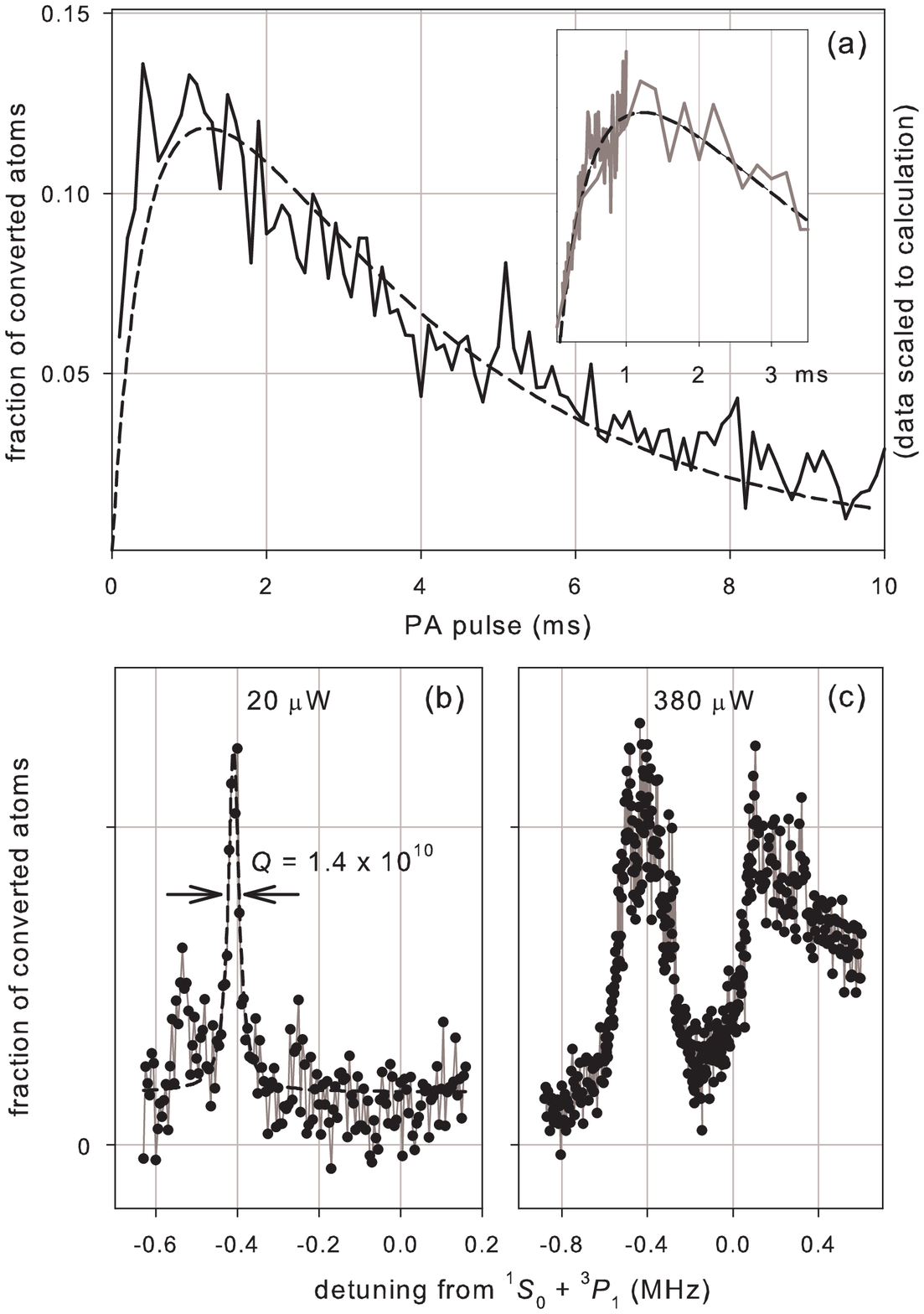}\hfill
\caption{(a) Solid line:  Fraction of the atom cloud detected after a molecule producing PA pulse followed by an atom clearing pulse and an atom recovery pulse.  The data are scaled to the calculation; the lowest bound on the peak atom-molecule conversion fraction is $3\%$.  Dashed line:  Calculated fraction of the atom cloud converted to $v=-2$ molecules after the PA pulse.  The curve parameters were measured independently, with the exception of the molecule lifetime, which was fixed at $3$ ms to best match the data.  The inset shows denser data in the first millisecond.  (b) Fraction of the atom cloud detected after a 1 ms PA pulse followed by the atom clearing and 20 $\mu$W atom recovery pulses, as a function of the L$_{\rm{BB}}$ frequency.  The sharp peak corresponds to the bound-bound transition from $v=-2$ to $v'=-1$.  Its fitted Lorentzian full width of 32(4) kHz corresponds to a quality factor $Q=1.4\times10^{10}$, the highest reported for a molecular vibronic transition.  (c) The same as (b), at the higher 380 $\mu$W atom recovery pulse power.  In this high power regime, atoms are also recovered via the $^1S_0+^3P_1$ thermal continuum, which is only 0.4 MHz away from $v'=-1$.}
\label{fig:MolCrLoss}
\end{figure}
}

\MolCrLossSchematic
\FCFmeasurements
\MolCrLoss
The rapid progress in laser cooling has given rise to many new fields of research.  One important example is the study of ultracold, dense clouds of molecules.  The molecules can exhibit new physical phenomena near quantum degeneracy \cite{YeCarrNJP09_ColdMolecules,MicheliNPhys06,YeOspelkausScience10_KRbReactions}.  For example, long-range anisotropic interactions are expected between heteronuclear polar molecules.  Such molecules have also been explored as a paradigm for quantum information and computation \cite{DeMillePRL02}.  On the other hand, homonuclear molecular dimers without a dipole moment present a metrological interest, for example in constraining the variation of the electron-proton mass ratio \cite{DeMillePRL08,ZelevinskyPRL08} or complementing atomic clocks by serving as time standards in the terahertz regime \cite{ZelevinskyPRL08}.  These molecules also provide an excellent testing ground for many possible approaches to creating large ultracold samples, trapping them to allow long interrogation times, and precisely controlling their quantum states.  The four promising routes toward trapped neutral molecules \cite{YeCarrNJP09_ColdMolecules} are direct control of polar molecule dynamics via electric fields; buffer gas cooling of magnetic species; direct laser cooling of a suitable class of molecules \cite{DeMilleShumanNature10_SrFLaserCooling}; and using magnetic or optical fields to combine laser cooled atoms into dimers \cite{DeMilleSagePRL05_RbCsViaPA,RomPRL04,LangPRL08_GroundStateRb2,NagerlDanzlNPhys10_Cs2GroundLattice,NiSci08}.  The latter process typically results in molecules with relatively small binding energies but nonetheless has played a major role in the study of new phenomena.  These dimers are also most promising for quantum control, since they are routinely produced at sub-$\mu$K temperatures.

In this Letter, we describe optical production of $^{88}$Sr$_2$ in the electronic ground state in an optical lattice.  In contrast to alkali-metal atoms, Sr atoms possess no electronic spin and cannot be combined into molecules via the magnetic Feshbach resonance technique \cite{ChinRMP10_FeshbachRes}.  However, Sr has the advantage of the intercombination (spin-forbidden) transition from the ground state, $^1S_0-^3P_1$ (7 kHz linewidth \cite{ZelevinskyPRL06}, 689 nm wavelength), which allows Doppler cooling to $<1$ $\mu$K \cite{KatoriPRL99} and provides several features that enable efficient photoassociation (PA) into molecules \cite{JonesRMP06,ZelevinskyPRL06}.  The unusually fast time scale (0.25 s) of ultracold molecule production is particularly important for establishing a short duty cycle in metrological applications.  Furthermore, there is active interest in exploring molecules of alkaline-earth-metal (and the isoelectronic Yb) atoms in various combinations with each other and with alkali-metal atoms, such as Sr$_2$, Yb$_2$, LiYb, RbYb, RbSr, and SrYb \cite{EscobarPRA08,YeBlattPRL11_SrOFR,TakahashiTakasuPRL12_Yb2Subradiant,TakahashiHaraPRL11_LiYb,GuptaHansenPRA11_LiYb,GorlitzNemitzPRA09_YbRbPA,MoszynskiTomzaPCCP11_LatticeSrYbTheory}.

Our approach takes advantage of several interesting properties presented by the narrow intercombination transition to achieve efficient transfer of atoms into molecules.  Figure \ref{fig:MolCrLossSchematic}(a) and its caption explain the notation used in this work.  The excited molecular potential dissociating to the $^1S_0+^3P_1$ atomic limit has a small attractive $C_3$ coefficient, with a $C_6$ coefficient similar to that of the ground state, resulting in relatively large wave function overlaps with the ground state van der Waals potential.  This leads to several Franck-Condon factors (FCFs) between high-lying vibrational states that are unusually close to unity.  We exploit these large FCFs for one-photon PA that is followed by spontaneous decay to predominantly a single vibrational level.  Another feature enabled by the narrow spectral line is the possibility to frequency resolve the least-bound electronically excited vibrational level with only a 440 $h\times$kHz binding energy corresponding to the dimer size exceeding 500 $a_0$ \cite{ZelevinskyPRL06} ($h$ is the Planck constant and $a_0\approx0.053$ nm is the Bohr radius).  This level has an optical length $l_{\rm{opt}}$ (the coupling strength proportional to the PA laser intensity and the free-bound wave function overlap) that is estimated to be $\sim10^4$ times larger than for the vibrational levels bound by hundreds of megahertz.  We tune the atom recovery laser to the bound-bound transition resonant with this level, and a significant fraction of subsequent spontaneous decays leads to a reappearance of ground state atoms.  Finally, this molecule creation method is the first step in a two-step sequence of producing Sr$_2$ in the absolute ground quantum state.
To this end, recent work \cite{MoszynskiSkomorowskiPRA12_Sr2Formation,MoszynskiSkomorowskiJCP12_Sr2Dynamics} predicts efficient two-photon coupling of the vibrational levels $v=-2,-3$ (Fig. \ref{fig:MolCrLossSchematic}) to the lowest level $v=0$ via an intermediate $v'\sim12$.
We have made the molecules in $v=-2$ via PA into $v'=-5$ (3.5 $h\times$GHz binding energy); furthermore, the calculated and measured FCF is also near unity for $(v,v')=(-3,-6)$, and our technique can be used for PA into $v'=-6$ (8.4 $h\times$GHz binding energy) by attaining a higher power output of the PA laser.
%Furthermore, theoretical results indicate that two-photon production of $^{88}$Sr$_2$ in the absolute ground quantum state from $v=-2$ should also be feasible, with an efficiency reduced by roughly $2\times$ \cite{MoszynskiPrivate}.
The ability to make deeply bound Sr$_2$ can enable a new class of highly precise molecular metrology tools \cite{ZelevinskyPRL08}.

Producing $^{88}$Sr$_2$ in the electronic ground state required performing precise two-photon PA spectroscopy.  The search for $v=-2$ and $v=-3$ was guided by the most accurate ground and excited state potentials available from Refs. \cite{TiemannSteinEPJD10_Sr2XPotential,TiemannSteinPRA08_Sr2XPotential}.  Starting with $5\times10^5$ 1 $\mu$K $^{88}$Sr atoms trapped in a 20 $\mu$K deep one-dimensional optical lattice formed at the 25 $\mu$m waist of a retroreflected 914 nm \cite{Ido3P1PRL03} laser beam (axial trap frequency 65 kHz; 95\% of atoms in the zero-point trap level), we carried out the spectroscopy via intermediate electronically excited levels, with varying detunings from $v'$.
The polarizations of the copropagating PA and detection lasers in this work are parallel to each other and to the small residual magnetic field at the atom trapping site and perpendicular to the lattice polarization; the lasers are aligned along the tight-trapping axis.
When the bound-bound laser (L$_{\rm{BB}}$) is near resonance, and the frequency of the free-bound laser (L$_{\rm{FB}}$, phase-locked to L$_{\rm{BB}}$ via a cavity-stabilized cooling laser) is scanned, Autler-Townes splitting of the PA resonance is observed via atom loss [Fig. \ref{fig:FCFmeasurements}(b), inset].  This splitting is equivalent to the generalized Rabi frequency of the bound-bound transition
\be
\Omega_{vv'}'=\sqrt{(\omega-\omega_0)^2+\Omega_{vv'}^2},
\label{eq:OmegaGen}
\ee
where $\omega$ is the frequency of L$_{\rm{BB}}$, $\omega_0$ corresponds to the bound-bound resonance, and $\Omega_{vv'}$ is the Rabi frequency.  Near the dissociation limit, $\Omega_{vv'}$ is directly related to the dimensionless FCF $f_{vv'}=|\langle v|v'\rangle|^2$, since $\Omega_{vv'}=\Omega_a\sqrt{2f_{vv'}}\alpha$, where $\alpha^2=1/3$ is the rotational line strength for $(J,J')=(0,1)$ transitions, and the atomic Rabi frequency is $\Omega_a=\Gamma_a\sqrt{s/2}$ in terms of the atomic spontaneous decay rate $\Gamma_a$ and saturation parameter $s$.

Different $(v,v')$ pairs yield complementary FCFs; the pairs relevant to this work are $(-2,-5)$ and $(-3,-6)$.  The corresponding FCFs were calculated to be 0.8(1), indicating a particularly strong coupling between the levels.  Figure \ref{fig:FCFmeasurements}(a) shows the Autler-Townes peak positions for $(v,v')=(-2,-5)$ at various frequencies of L$_{\rm{BB}}$.  The difference between the two curve branches is $\Omega_{(-2,-5)}'$ and is fitted to Eq. (\ref{eq:OmegaGen}).  The FCF is found to be $f_{(-2,-5)}=1.1$, which is consistent with 0.8(1) if the 27 $\mu$m PA laser beam waist measurement has an error of just 15\%.  In contrast, Fig. \ref{fig:FCFmeasurements}(b) shows the Autler-Townes peak positions and fit for $(v,v')=(-3,-4)$.  With $60\times$ more L$_{\rm{BB}}$ power, the splitting is $4.4\times$ smaller, leading to the experimentally determined ratio $f_{(-3,-4)}/f_{(-2,-5)}=0.9\times10^{-3}$.  This agrees well with the calculated ratio of $1.1\times10^{-3}$.  While the $(v,v')=(-3,-6)$ pair was not used in this work, our measurements have confirmed that its FCF is also near unity.  Note that the fits such as shown in Fig. \ref{fig:FCFmeasurements} yield the binding energies for $v=-2$ and $v=-3$ levels $(J=0)$ as 1400.1(2)$h\times$MHz and 5110.6(2)$h\times$MHz, which agree with our calculations to 1.3\% and 0.3\%.

The FCFs between weakly bound vibrational levels were computed from adiabatic single-channel potentials fitted to high resolution Fourier transform spectra of Refs. \cite{TiemannSteinPRA08_Sr2XPotential,TiemannSteinEPJD10_Sr2XPotential}.  They were also calculated in Refs. \cite{MoszynskiSkomorowskiJCP12_Sr2Dynamics,MoszynskiSkomorowskiPRA12_Sr2Formation} by describing the rovibrational dynamics in the excited
electronic states by a coupled multichannel fully nonadiabatic Hamiltonian.  The large calculated value of $f_{(-2,-5)}=0.8(1)$ is surprising at first glance,
since the wave function of $v'=-5$ shows 38 oscillations, while
the wave function of $v=-2$ shows 61.  However, according to the Franck-Condon
principle \cite{CondonPR28}, molecular transitions most readily occur between rovibrational levels
of two different electronic states with nearly the same classical turning points.  In the adiabatic picture the classical turning points of $v'=-5$ and $v=-2$ are very close, with their respective inner points at 7.1 and 7.5$a_0$ and
outer points at 47.1 and 50.0$a_0$.  This near-coincidence explains the exceptionally large overlap between the
two wave functions at large interatomic distances, as plotted in Fig. \ref{fig:MolCrLossSchematic}(b).  A similarly large
overlap is predicted for $v'=-6$ and $v=-3$, with the respective outer turning points at 40.5 and 40.1$a_0$ \cite{MoszynskiPrivate}.

The $^{88}$Sr$_2$ molecule formation relies on one-photon PA into $v'=-5$ followed by spontaneous decay into $v=-2$.
The PA induced atom density loss from the lattice is described by $\dot{n}_a=-2Kn_a^2$,
where $K$ is the two-body loss rate and we have neglected the finite lifetime of the trapped atoms ($\sim10$ s).  Thus $n_a(t)=n_0/(1+2Kn_0t)$, where $n_0$ is the initial atom density.  Hence the time evolution of the molecule density is approximately given by $\dot{n}_m\approx0.3f_{vv'}Kn_a^2-n_m/\tau_c$, where $\tau_c=(n_0\gamma_c)^{-1}$ is the lifetime parameter for the created molecules assuming similar rates for molecule-atom and molecule-molecule collisional losses, and the 0.3 factor results from the spontaneous decay branching ratio from $J'=1$ to $J=0$ \cite{KotochigovaPRA09}.  (Note that with a simple repumping scheme out of $J=2$, a tripled atom-molecule conversion efficiency is expected.)  Therefore, the created molecule density as a fraction of $n_0$ can be expressed as
\be
\frac{d}{dt}\left(\frac{n_m}{n_0}\right)=\frac{b}{(1+at)^2}-c\left(\frac{n_m}{n_0}\right).
\label{eq:MolDensEvol}
\ee
For PA to $v'=-5$, the rates $a=2Kn_0$ (and therefore $b=0.3f_{vv'}Kn_0$) were determined by measuring atom loss as a function of the L$_{\rm{FB}}$ pulse length.  At the L$_{\rm{FB}}$ power of 3.2 mW, we find $a=1680$/s and $b=220$/s.  The loss rate $c$ also has to be experimentally determined.  If molecule-atom collisions dominate, it can be estimated as $c\approx2hn_0R_6/\mu\sim200$/s \cite{JuliennePCCP11_UniversalCollisions} for our densities $n_0\sim2\times10^{12}$/cm$^3$, where $\mu$ is the molecule-atom reduced mass; the van der Waals length $R_6=0.5(2\mu C_6/\hbar^2)^{1/4}\approx5.0$ nm, where $\hbar=h/(2\pi)$ and $C_6$ is the relevant attractive coefficient.  This $\sim5$ ms estimated collisional lifetime necessitates forming molecules on the millisecond time scale.  While it requires relatively large intensities of L$_{\rm{FB}}$, the undesirable photon scattering by the atomic transition is suppressed due to the narrow line.  The numerical solution $(2n_m/n_0)$ to Eq. (\ref{eq:MolDensEvol}) is plotted in Fig. \ref{fig:MolCrLoss}(a) for various PA pulse durations.  As expected, for a sufficiently large PA rate $a$, the number of photoassociated atoms first rapidly grows and then gradually falls as the production slows relative to collisional loss.

After the PA pulse is applied to the atoms, they are exposed to a low-intensity clearing pulse resonant with the strong $^1S_0-^1P_1$ transition at 461 nm.  The 0.2 ms pulse length is twice the minimum duration needed to remove the nonphotoassociated atoms to below the background level of the imaging setup; no systematic dependence on longer clearing pulse durations is observed.  Then a 0.1 ms atom recovery pulse is applied to the molecules created in $v=-2$.  This pulse is resonant with $(v,v')=(-2,-1)$; the calculated $f_{(-2,-1)}=1.4\times10^{-5}$.  As discussed earlier, $v'=-1$ has an exceptionally strong coupling to the free atom state:  $l_{\rm{opt}}\sim8\times10^5$ $a_0/$(W/cm$^2$) from calculations and $\sim4\times10^5$ $a_0/$(W/cm$^2$) from previous measurements \cite{ZelevinskyPRL06}.  Because of the small binding energy and large $l_{\rm{opt}}$, a significant fraction ($30\%$) of the $v'=-1$ molecules is expected to spontaneously decay into ground state atoms that get recaptured in our optical lattice.

Figure \ref{fig:MolCrLoss}(a) shows the result of the $^{88}$Sr$_2$ electronic ground state molecule creation followed by conversion of the molecules back to atoms.  An overall scaling factor has been applied to the data.  The expected curve is obtained from Eq. (\ref{eq:MolDensEvol}), with the molecule lifetime chosen to best match the data.  This results in $c\sim380/$s, or a $3$ ms lifetime of the $v=-2$ molecules.  The calculated curve predicts that at the given L$_{\rm{FB}}$ power, 12\% of the atoms should be converted to molecules at optimal PA pulse lengths of 1 ms.  The actual fraction of recovered atoms is 1\%, which corresponds to a conversion of $3\%$ of the atoms into molecules assuming a $30\%$ efficiency of the recovery pulse.  Given this uncertainty, we conclude that we create $2(1)\times10^4$ molecules with the 1 ms PA pulse.  The lower-than-expected number of recovered atoms is likely due to a reduced efficiency of the atom recovery step rather than of molecule formation, since recapture depends on the lattice dynamics of atoms with a wide kinetic energy spectrum.  Note that no significant heating of the recovered atoms is observed at the end of the process.  Detecting $v=-2$ molecules produced in the $J=2$ state could yield more information on the overall efficiency.  Furthermore, preliminary work indicates that purified $^{88}$Sr$_2$ samples can have very long lifetimes on the 10 ms and 1 s scales in one- and three-dimensional lattices, respectively, limited by bimolecular collisions.  Detailed studies of molecule losses at these extended times in the regime of few molecules per lattice site could help ascertain the efficiency factors.
%We project that as a three-dimensional optical lattice is imposed on the atoms after the 1 ms PA pulse and the atom clearing pulse are applied, the molecules can be confined on time scales exceeding 1 s \cite{NagerlDanzlNPhys10_Cs2GroundLattice,YeChotiaPRL12_20sKRb3DLattice}.

Atom recovery via the least-bound excited molecular level is confirmed by choosing the optimal 1 ms molecule producing pulse duration and measuring the recovered atom fraction as a function of the L$_{\rm{BB}}$ frequency, as shown in Fig. \ref{fig:MolCrLoss}(b).  The narrow peak corresponds to the bound-bound transition from $v=-2$ to $v'=-1$.  Its fitted Lorentzian full width of 32(4) kHz, approximately twice the expected natural width, corresponds to a high quality factor $Q=1.4\times10^{10}$ of the $^{88}$Sr$_2$ molecular vibronic transition.  At stronger saturation, the resonance broadens as in Fig. \ref{fig:MolCrLoss}(c), and the atoms can also be recovered via the electronically excited thermal continuum, which is 0.4 MHz blue-detuned from the molecular line.

In conclusion, we have produced stable ultracold lattice-confined $^{88}$Sr$_2$ molecules in the electronic ground state that are bound by 0.05 cm$^{-1}$.  Our technique is the first step of a two-step projected sequence for achieving large samples of ultracold Sr$_2$ in the absolute ground quantum state.  This has immediate applications in precise time and frequency metrology, studies of fundamental constant variations, and investigations of chemical reactions at ultralow kinetic energies and highly nonthermal internal state distributions.  The all-optical molecule creation scheme relies on a large wave function overlap between a pair of electronically excited and ground rovibrational levels that is characteristic of the alkaline-earth-metal-atom system.  The entire experimental trajectory ending in the $\mu$K molecule sample takes only 0.25 s.  The atom recovery scheme, proving the creation of molecules, relies on one-photon excitation to the least-bound vibrational level near the intercombination line dissociation limit.  This level can be resolved due to the narrow optical line width and has an unusually large coupling to the thermal continuum of the ground electronic state.  To achieve the goals of this work, we have performed one- and two-photon spectroscopic studies of $^{88}$Sr$_2$.  Note that related results are reported for $^{84}$Sr in a Mott-insulator state \cite{SchreckStellmerArxiv12_Sr2}.

We acknowledge invaluable contributions to theoretical studies of our system by R. Moszynski, C. Koch, and W. Skomorowski, and thank them for the data in Fig. \ref{fig:MolCrLossSchematic}(b).  We also thank P. Julienne and B. H. McGuyer for many fruitful discussions and K. Bega for contributions in the earlier stages of the experiment.  This work was partially supported by the ARO, the AFOSR, and the NSF.

%\begin{figure}[!h]
%    \centering
%    \includegraphics[trim = 0in 0in 0in 0.5in, clip, width=3.4in, angle=0]{Figs/8400MHz5100MHzAutlerTownesPeaks.pdf}
%    \caption{Positions of Autler-Townes peaks that emerge when the bound-bound PA2 laser is fixed near a molecular resonance while the free-bound PA1 laser frequency is scanned.  The dashed line is the measured position of the one-photon photoassociation peak near $v'=-6\;(J=1)$, with the $-8,430$ MHz binding energy \cite{ZelevinskyPRL06}; the two-photon transition is to the $v=-3\;(J=0)$ level.}
%    \label{fig:8400MHz5100MHzAutlerTownesPeaks}
%\end{figure}

\end{document}